\documentclass[aps,prd,twocolumn,
amssymb,amsmath,showpacs,a4paper,superscriptaddress,nofootinbib,longbibliography]{revtex4-2}
\usepackage{graphicx}
\usepackage{braket}
\usepackage{amsfonts}
\usepackage{amsmath}
\usepackage{bbold}
\usepackage{hyperref}
\usepackage{adjustbox}
\usepackage{float}
\usepackage[nameinlink]{cleveref}
\usepackage[table,xcdraw]{xcolor}
\usepackage{mathrsfs} 
\newcommand{\beq}{\begin{equation}}
\newcommand{\eeq}{\end{equation}}
\newcommand{\beqn}{\begin{eqnarray}}
\newcommand{\eeqn}{\end{eqnarray}}
\newcommand{\beqs}{\begin{subeqnarray}}
\newcommand{\eeqs}{\end{subeqnarray}}

\definecolor{darkblue}{RGB}{16,78,139}

\begin{document}
	\preprint{APS/123-QED}
	\title{Black hole spectral instabilities in the laboratory: Shallow water analogue}
	
	\author{Mateus Malato Corrêa}
	\email{malato.mateus@gmail.com}\affiliation{
		Programa de P\'{o}s-Gradua\c{c}\~{a}o em F\'{i}sica, Universidade Federal do Par\'{a}, 66075-110, Bel\'{e}m, PA, Brazil
	}
        \affiliation{Center of Gravity, Niels Bohr Institute, Niels Bohr Institute, Blegdamsvej 17, 2100 Copenhagen, Denmark}
	\author{Caio F. B. Macedo}
	\email{caiomacedo@ufpa.br}
	\affiliation{Faculdade de F\'{i}sica, Campus Salin\'{o}polis, Universidade Federal do Par\'{a}, 68721-000, Salin\'{o}polis, Par\'{a}, Brazil
        }
	\author{Rodrigo Panosso Macedo}
	\email{rodrigo.macedo@nbi.ku.dk}
    \affiliation{Center of Gravity, Niels Bohr Institute, Niels Bohr Institute, Blegdamsvej 17, 2100 Copenhagen, Denmark}
	\author{Leandro A. Oliveira}
	\email{laoliveira@ufpa.br}
	\affiliation{Faculdade de F\'{i}sica, Campus Salin\'{o}polis, Universidade Federal do Par\'{a}, 68721-000, Salin\'{o}polis, Par\'{a}, Brazil
	}
	\date{\today}
	
	\begin{abstract}

Small deviations in the spacetime around black holes can lead to instabilities in the underlying quasinormal mode spectrum, potentially altering the hierarchy of its overtones. A practical way to induce such spectral instability is by introducing small modifications to the effective potential governing the dynamics of fluctuations in the black hole spacetime. While finding a physically meaningful interpretation for such ad hoc modifications in an astrophysical context can be challenging, analogue black hole models provide an alternative framework to explore their effects and study the instabilities. In this work, we consider an analogue black hole modeled by a draining bathtub flow and demonstrate that vorticities in the fluid introduce a small bump in the effective potential of the wave equation. This naturally realizes a physically motivated version of the elephant and the flea configuration. We analyze the spectrum using two complementary approaches: direct mode computation via two distinct frequency-domain methods and time evolution of initial perturbations. As in astrophysical black holes, the vorticities destabilizes the QNM spectrum of the analogue system, possibly yielding time evolution with long-lived ringing effects, akin to those observed for massive fields in curved spacetimes.

	\end{abstract}
	\maketitle
	\section{\label{sec:Introduction}Introduction}
       The coalescence of two black holes is a highly energetic and dissipative event. According to General Relativity, the remnant object is a black hole (BH) described solely by three parameters: mass, charge, and angular momentum \cite{Chandrasekhar:1985kt}. The post-merger dynamics is characterized by a ringdown phase, during which the spacetime undergoes a relaxation process toward stationarity. In this regime, gravitational waves (GWs) are represented by small spacetime fluctuations that oscillate and decay exponentially as energy flows into the black hole and propagates toward the wave zone. The characteristic frequencies modulating part of this ringdown are called quasinormal modes (QNMs) and are directly related to the structure of the spacetime \cite{Nollert;1999,Kokkotas:1999bd,Berti;2009,Zhidenko;2011}.
        
These frequencies have been computed for various configurations within and beyond General Relativity—see e.g.~\cite{Regge:1957td, Zerilli:1970wzz,Kokkotas:1988fm, Teukolsky:1972my,Blazquez-Salcedo:2016enn, Molina:2010fb,McManus:2019ulj, Macedo:2016wgh} for a non-exhaustive list—and detecting them allows for fundamental tests of gravitational theory~\cite{Kokkotas:1999bd,Berti;2009,Zhidenko;2011,Barausse:2014tra,Dreyer:2003bv,Berti:2005ys}. Observations of the GWs emitted during binary black hole mergers by the LIGO-Virgo-KAGRA Collaboration have provided strong evidence for the presence of the fundamental mode in the ringdown phase \cite{LIGOScientific:2016lio, KAGRA:2021vkt, LIGOScientific:2021sio}. Future improvements in detector sensitivity, the development of next-generation Earth-based observatories, and the introduction of space-based detectors such as LISA will open a new window for detecting overtones and extreme mass-ratio inspirals (EMRIs), offering novel opportunities to test General Relativity \cite{Barausse:2020rsu}.

Due to the dissipative nature of black hole perturbation theory—placing it within the realm of non-Hermitian physics~\cite{Ashida:2020dkc,Jaramillo:2020tuu}—the QNM spectrum is, in principle, highly sensitive to small deviations in the system. Motivated by both formal aspects of the QNM problem and the influence of the astrophysical environment surrounding black holes, it has been shown that \textit{small deviations} in the system can lead to \textit{significant} changes in the frequency spectrum. Early evidence of QNM spectral instabilities was reported in \cite{PhysRevD.53.4397,nollert2,Aguirregabiria:1996zy,Vishveshwara:1996jgz}, and more recently, this problem has been addressed using the pseudospectrum approach \cite{Jaramillo:2020tuu, Jaramillo:2021tmt}. Pseudospectrum analysis has now revealed QNM instabilities in various spacetimes~\cite{Destounis:2021lum,Boyanov:2022ark,Sarkar:2023rhp,Destounis:2023nmb,Arean:2023ejh,Cownden:2023dam,Boyanov:2023qqf,Cao:2024oud,Luo:2024dxl,Chen:2024mon,Cai:2025irl,Tobias2025}.

A common approach to inducing spectral instabilities in practice is to introduce small \textit{ad hoc} modifications to the effective potential governing wave propagation in a black hole spacetime. While it is well understood that higher overtones are more susceptible to destabilization~\cite{Vishveshwara:1996jgz,Jaramillo:2020tuu}, certain modifications that alter the potential at large distances can even significantly affect the fundamental (slowest-decaying) mode~\cite{PhysRevD.53.4397,Jaramillo:2020tuu,Cheung:2021bol}. However, an open challenge remains in understanding the physical reality of such modifications~\cite{Boyanov:2024fgc,Cardoso:2024mrw}, their impact on wave signals, and their detectability \cite{Jaramillo:2021tmt,Spieksma:2024voy}.
  
A notable example of such a destabilization is the ``elephant and the flea” configuration~\cite{Cheung:2021bol}, in which a small bump is added to the effective potential at a relatively large distance from its original peak. This \textit{ad hoc} perturbation is often interpreted as representing mass surrounding the black hole, leading to the conclusion that environmental effects can drastically alter the QNM spectrum \cite{Destounis:2023ruj,Cardoso:2024mrw,Cheung:2021bol}. When extracted from time-domain profiles, this frequency spectrum does not exhibit instability at early times but shows modifications at intermediate times and in the late-time signal tail. These changes introduce oscillatory features resembling trapped modes, echoes, and long-lived modes \cite{Cardoso:2020nst,Destounis:2023ruj,Correa:2024xki}. The relationship between QNM frequencies and time-domain behavior can also be analyzed by studying the Green function poles and their role in wave scattering~\cite{PhysRevD.110.084018}.

While modifications in wave equation's potential are still often considered toy models in astrophysical contexts, analogue black holes offer a compelling alternative to study QNM instabilities from first principles. The behavior of small disturbances in black hole spacetimes—such as those occurring during ringdown—can, to some extent, be replicated by gravity waves propagating in a fluid flow \cite{Schutzhold:2002rf,patrick2020analogyblackholesbathtub,PhysRevLett.121.061101}. This is possible because certain fluid configurations mimic the behavior of an event horizon, exhibiting kinematic features similar to those found in black hole spacetimes. This resemblance has led to the study of phenomena such as Hawking radiation in analogue systems, reinforcing the connection between these so-called analogue spacetimes and real black holes \cite{Unruh:1980cg,Barcelo:2005fc,Cardoso:2013jus, Visser:1997ux, Novello:2002qg}.

Research on these fluid analogues has revealed results similar to those in black hole spacetimes, including absorption, scattering, QNMs, quasibound states, and superradiance \cite{Benone:2014nla,Benone:2015jda, Benone:2018xct, Oliveira:2020hal,Vieira:2025ljl,Oliveira:2024quw}. The experimental feasibility of creating such systems in the laboratory has enabled the detection of QNMs \cite{Torres:2020tzs}, the observation of superradiance in scattering waves \cite{Torres:2016iee}, and empirical confirmation of QNM enhancement in confined systems \cite{Smaniotto:2025hqm}. Furthermore, pseudospectrum analysis of analogue black holes has also indicated the presence of spectral instabilities \cite{Tobias2025}.

This work contributes to the ongoing efforts to understand QNM instability by considering an analogue ``elephant and the flea” configuration. In this context, the modification of the potential is more natural, as it arises not from an \textit{ad hoc} alteration of the wave equation but rather as a direct consequence of the system’s fluid dynamics. Specifically, we show that a vortex in the fluid flow, modeled by a solution of the Navier-Stokes equation \cite{BURGERS1948171, Rott1958ZaMP}, naturally introduces a small bump in the effective potential. We then explore the consequences of this configuration for the QNM spectrum.

The paper is organized as follows: In Section~\ref{sec:gravity}, we derive the equations for gravity waves in a fluid flow with vorticity and describe how they can be formulated in terms of an effective spacetime. We also introduce the configuration in which we study frequency spectrum destabilization and incorporate the Burgers-Rott vortex to model the ``bump” in the potential. Section~\ref{sec:timedomain} presents the decomposition of the scalar field describing gravity waves in this system and the procedure for obtaining the time evolution of perturbations. In Section~\ref{sec:Spectrum}, we outline the conventions used to track mode migration and discuss the perturbation’s impact on QNMs. Section~\ref{sec:results} details our findings, highlighting spectral migration and destabilization, as well as time-domain evolution. Finally, we summarize our conclusions in Section~\ref{sec:conclusion}.

	\section{Gravity waves in a fluid with vorticity}\label{sec:gravity}
	Let us start by reviewing the spacetime analogy. Considering a two-dimensional incompressible inviscid fluid with vorticity, we write the following differential equations for linear perturbations to the background fluid flow~\cite{patrick2020analogyblackholesbathtub, PhysRevLett.121.061101} 
	\beq
	\frac{\partial \vec{v}}{\partial t}+\left(\vec{V}\cdot \nabla\right)  \vec{v}+ \left(\vec{v}\cdot \nabla\right) \vec{V}+g_{\rm \ell} \nabla h=0,
	\label{euler}
	\eeq
	and
	\beq
	\frac{\partial h}{\partial t}+\left(\vec{V}\cdot \nabla\right)h+H\nabla\cdot \vec{v}=0,
	\label{cont}
	\eeq
	where $\vec{V}$ is the background velocity of the fluid, $\vec{v}$ is the perturbation velocity, $H$ is the height of the non-perturbed fluid, $h$ is a small displacement of the height of the fluid and $g_{\rm \ell}$ is the local gravity.
	
	We now chose the fluid properties in such a way that we describe a BH analogue. Let us first notice that we can split the velocity in the following way
	\beq
	\vec{V}=\vec{V}_{\rm irr}+\vec{V}_{\rm rot},
	\label{bvel}
	\eeq
	where $\vec{V}_{\rm irr}$  ($\vec{V}_{\rm rot}$) is the irrotational (rotational) part of the background flow. For the fluid flow, we describe the background vorticity as
	\beq
	\vec{\Omega}=\nabla\times\vec{V}_{\rm rot}=\Omega \widehat{k},
	\label{bvort}
	\eeq
	where $\Omega=|\vec{\Omega}|$.
	The perturbation velocity is represented by a Helmholtz decomposition, as
	\beq
	\vec{v}=\nabla \phi+\widetilde{\nabla} \psi,
	\label{pvel}
	\eeq
	where $\widetilde{\nabla}=\widehat{k}\times\nabla$ is called cogradient operator, satisfying~$\left(\nabla\cdot\widetilde{\nabla}\right)\psi=0$.
    
	Finally, considering Eqs.~\eqref{euler},~\eqref{cont} and~\eqref{pvel}, we obtain the following system of equations
	\beqn
	&&\frac{\partial \phi}{\partial t}+\left(\vec{V}\cdot \nabla\right)\phi-\psi\Omega +g h=0, \label{euler1}\\
	&&\frac{\partial \psi}{\partial t}+\left(\vec{V}\cdot \nabla\right) \psi+\phi\Omega=0,\label{vort1}\\
	&&\frac{\partial h}{\partial t}+\left(\vec{V}\cdot \nabla\right)h+H\nabla^2\phi=0. \label{cont1}
	\eeqn
	Eqs.~\eqref{euler1} and~\eqref{vort1} are exact in two particular cases only: solid body rotation ($\Omega=constant$) and irrotational flow ($\Omega=0$). Then, other rotational flow regimes, Eqs.~\eqref{euler1} and~\eqref{vort1} provide an approximate description of gravity waves in a rotational flow~\cite{patrick2020analogyblackholesbathtub, PhysRevLett.121.061101}. 
	
	Substituting $h$ from Eq.~\eqref{euler1} in Eq.~\eqref{cont1}, considering Eq.~\eqref{vort1} and that background vorticity satisfies $$\frac{\partial \Omega}{\partial t}+\left(\vec{V}\cdot \nabla\right)\Omega =0,$$ we obtain
	\beqn
	&&\frac{\partial}{\partial t}\left[ -\frac{\partial \phi}{\partial t}-\left(\vec{V}\cdot \nabla\right)\phi\right]+\left(\vec{V}\cdot \nabla\right)\left[-\frac{\partial \phi}{\partial t}-\left(\vec{V}\cdot \nabla\right)\phi\right] \nonumber\\
	&& + c^2\nabla^2\phi-\Omega^2 \phi=0,
	\label{waveeq}
	\eeqn
	where $c=\sqrt{gH}$ is the speed of the gravity waves. With the above we can make an analogy between the description of gravity waves and general relativity, we may rewrite Eq.~\eqref{waveeq} in the form
	\beq
	\frac{\partial \chi^\mu}{\partial x^\mu}-\Omega^2 \phi=0,
	\label{waveeq1}
	\eeq
	where $\chi^\mu=\left(\chi^0,\vec{\chi}\right)$ and $\mu=0,1,2,3$, being
	\beqn
	&&\chi^0= -\frac{\partial \phi}{\partial t}-\left(\vec{V}\cdot \nabla\right)\phi,\\
	&&\vec{\chi}= \left[-\frac{\partial \phi}{\partial t}-\left(\vec{V}\cdot \nabla\right)\phi\right]\vec{V}+c^2\nabla\phi.
	\eeqn
	Furthermore, Eq.~\eqref{waveeq1} may be written as
	\beq
	\frac{1}{\sqrt{-g}}\frac{\partial }{\partial x^\mu}\left(\sqrt{-g} g^{\mu\nu}\frac{\partial \phi}{\partial x^\nu}\right)-\frac{\Omega^2 }{c^2}\phi=0,
	\label{waveeq2}
	\eeq
	where $g^{\mu\nu}$ is the contravariant metric and $g$ is the determinant of the covariant metric.
	
	With the above we have the analogy constructed through the metric $g_{\mu\nu}$. We can represent the line element in polar coordinates as
	\beq
	ds^2=-\left(c^2-|\vec{V}|^2\right)dt^2-2 V_r dt dr-2 r V_\theta dt d\theta+dr^2 + r^2 d\theta^2,
	\label{line}
	\eeq
	where $t$ is the time coordinate, $r$ is the radial coordinate, $\theta$ is the angular coordinate, $V_r$ is the radial component of the background velocity and $V_\theta$ is the angular component of the background velocity.
	We rewrite the line element~\eqref{line} using the following coordinate transform
	\beqn
	&&dt\rightarrow dt-\frac{V_r}{c^2 f(r)}dr,\\
	&&d\theta \rightarrow d\theta -\frac{V_r\,V_\theta}{r c^2 f(r)}dr,
	\eeqn
	where $f(r)=1-\dfrac{V_r^2}{c^2}$. We obtain
	\beq
	ds^2=-\left(c^2-|\vec{V}|^2\right)dt^2-2 r V_\theta dt d\theta+f(r)^{-1}dr^2 + r^2 d\theta^2.
	\label{line1}
	\eeq
        Assuming incompressibility, i.e., $\nabla\cdot \vec{V}=0$, we may write the radial background velocity as
	\beq
	V_r=-c \frac{r_{\rm h}}{r},
	\label{rad_vel}
	\eeq
	where $r_{\rm h}$ is the analogue event horizon radius. The angular background velocity is such that $\left|{V_\theta}/{V_r}\right|\ll1$, but there is sufficient background vorticity to play the role of effective mass of the field $\phi$, namely
	\beq
	\Omega=\frac{1}{r}\frac{\partial}{\partial r}\left(rV_\theta\right).
	\eeq
	
	In this paper, we use a model based on a Burgers-Rott vortex \cite{BURGERS1948171,Rott1958ZaMP}, represented by a Gaussian function, as follows
	\beq
	\Omega=\frac{\Gamma}{2\pi \sigma_{\rm v}^2}\exp\left[-\frac{\left(r-r_0\right)^2}{2\sigma_{\rm v}^2}\right],
    \label{eq:burgers}
	\eeq
	where $\Gamma$ is the circulation at radial infinity, $\sigma_{\rm v}\equiv \sqrt{{\mu_{\rm k}}/{\alpha}}$ is the width of the vortex, $\alpha$ is the radial draining at the center of the vortex, $\mu_{\rm k}$ is the kinematic viscosity of the vortex and $r_0$ is the position of the center of the vortex.
	
	We consider that the effective spacetime of this system of gravity waves is produced from a radial draining of the fluid at the $r=0$. Then, the Burgers-Rott vortex placed at $r=r_0$ affects the perturbation velocity only, acting like an effective mass of the field $\phi$. Summarizing, considering the line element~\eqref{line1} and Eq.~\eqref{rad_vel}, we have that the effective spacetime is described by 
	\beq
	ds^2=-f(r)\,c^2dt^2+f(r)^{-1}dr^2 + r^2 d\theta^2,
	\label{line2}
	\eeq
	where $f(r)=1-\dfrac{r_{\rm h}^2}{r^2}$. 
	
	\section{Perturbations and wave equation decomposition}\label{sec:timedomain}
	
	Considering the line element~\eqref{line2} and Eq.~\eqref{waveeq2}, we can decompose the field perturbation as
	\beq
	\phi(t,r,\theta)=\frac{1}{\sqrt{r}}\sum_{m=-\infty}^{m=\infty}\Phi(t,r)\exp\left(i\,m\,\theta\right),
	\eeq
	where the summation over integer $m$ is found by requiring angular periodicity on $\phi$. The function $\Phi(t,r)$ obeys the following partial differential equation
	\beq
	\left[-\frac{\partial^2}{\partial t^2}+\frac{\partial^2}{\partial x^2}-V_{\Gamma}(r)\right]\Phi(t,x)=0,
	\label{waveeq3}
	\eeq
	with $x$ being the tortoise coordinate, given by
	\beq\label{EQ:tortoise}
	x=\frac{r}{c}+\frac{r_{\rm h}}{2c}\ln\left(\frac{r-r_{\rm h}}{r+r_{\rm h}}\right),
	\eeq
	and $V_{\Gamma}(r)$ is the effective potential, given by
	\beq
	V_{\Gamma}(r)=f(r)\left[c^2\left(\frac{m^2-1/4}{r^{2}}+\frac{5r_{\rm h}^2}{4r^4}\right)+\Omega^2\right].
	\eeq

\begin{figure}[htpb!]
\centering
\includegraphics[width=0.45\textwidth]{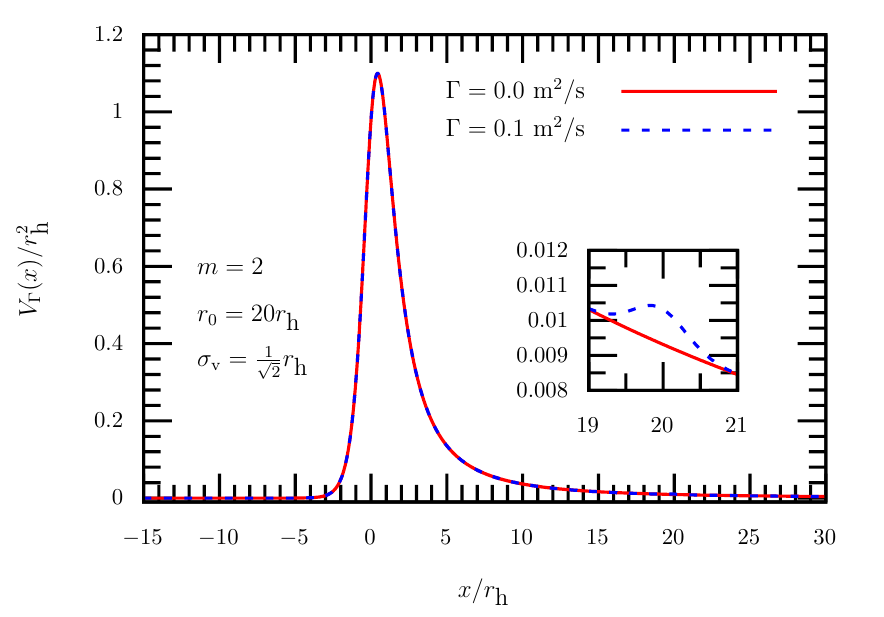}

\caption{Effective potential $V_\Gamma(x)$, as a function of the Regge-Wheeler coordinate $x$, for $m=2$ and $r_0 = 20\,r_{h}$. The small bump highlighted in the inset arises naturally as the fluid vorticity in this configuration for black hole analogues.}
\label{pot}
\end{figure}

In Fig.~\ref{pot}, we plot the effective potential $V_{\Gamma}(x)$, as a function of the tortoise, coordinate $x$. We include the Burgers-Rott vortex described by Eq.~\eqref{eq:burgers}, selecting parameters such that the background flow is still the primary driver for the black hole analogue metric. 

One observes that the potential shares the main features of the one studied in the ``elephant and the flea" spectral stability analysis in the Schwarzschild spacetime ~\cite{Cheung:2021bol}. Despite the similar model, we emphasise that in the astrophysical scenario, a small bump is an {\em ad hoc} contribution added to the wave equation's potential. Here, the tiny bump arises naturally from the analogue model's equation, providing us with a top-down approach to  realize a physically motivated version of the elephant and the flea configuration.
    
\subsection{Time evolution of initial data}

One way to observe the effects of the additional bump in the potential is to look into the evolution of initial data subjected to Eq.~\eqref{waveeq3}. In this work, we employ the method of lines as follows. 

\begin{enumerate}
    \item We discretize the radial coordinate $r \rightarrow r_{j}$, tortoise coordinate $x \rightarrow x_{j}$, the wave function $\Phi(t, x) \rightarrow \Phi_j(t),$ and second-order spatial derivative, namely, 
\begin{equation}
\dfrac{\partial^2  \Phi(t, x)}{\partial x^{2}} \rightarrow \dfrac{1}{\Delta x^2}\left[\Phi_{j+1}(t)-2\Phi_{j}(t)+\Phi_{j-1}(t)\right] +{\cal{O}}(\Delta x^2),    
\end{equation}
where $\Delta x$ is the (uniform) grid spacing. 
    \item By employing the discretizations and using the definition $\zeta_{j}(t) \equiv \frac{d \Phi_{j}(t)}{dt}$, we obtain
\begin{align}
&\dfrac{d \Phi_{j}}{dt}=\zeta_{j},
\label{ode1}\\
&\dfrac{d \zeta_{j}}{dt}=\frac{1}{\Delta x^2}\left(\Phi_{j+1}-2\Phi_{j}+\Phi_{j-1}\right) - V_\Gamma(r_j)\,\Phi_{j},
\label{ode2}
\end{align}
which forms a set of coupled ordinary differential equations.
    \item We solve the set~\eqref{ode1} and \eqref{ode2} with a fourth-order Runge-Kutta method subjected to the following initial conditions
    \begin{align}
&\Phi_j(t=0) = \exp\left[ \frac{-\left(x_j -x_{\rm p}\right)^2}{2 \sigma_{\rm p}^2}\right]\quad\textrm{and}\\
&\zeta_{j}(t=0)= 0,
\label{Gauss}
\end{align}
where $x_{\rm p}$ is the position of the center of the peak of the Gaussian function (middle point), and $\sigma_{\rm p}$ sets its width. With this we cover the solution for the whole space (given a point $j$ of the grid) in the time range defined by the integration.
\end{enumerate}

By using the above, we evolve the wave function subjected to the initial conditions and extract it at some external ``observer'' point, which we adopt to be $r_{\rm ext}=25r_h$ (with associated tortoise coordinate $x_{\rm ext}$). The numerical grid has outer boundaries satisfying $x_{\rm bound} \gg x_{\rm ext}$, and we extract the GW within the initial data's  future causal domain, i.e. within a time interval not contaminated by the noise emanating from the numerical boundaries. We use for the initial condition $x_{\rm p} = \sqrt{2} \,r_{\rm h}$, $\sigma_{\rm p} = r_{\rm h}^2$ and typically set $\Delta x=0.01$, but checking that our results are accurate enough by testing different grid sizes in some simulations. 

In a clean background, i.e., without the bump, the ringdown stage is relatively clear, presenting a superposition of damped sinusoids given by the quasinormal modes frequencies. In sec.~\ref{sec:results} we investigate how small perturbations affects this picture.

	\section{Spectrum of Quasinormal modes}\label{sec:Spectrum}
        To work in the frequency domain, we assume a Fourier decomposition $\Phi = e^{-i \omega t}\psi(r)$ in Eq.~(\ref{waveeq3}),
        \begin{equation}
            \frac{d^{2}\psi}{dx^{2}}+\left(\omega^{2}-V_{\Gamma}(r)\right)\psi=0.
        \end{equation}
        The quasinormal modes are described by the characteristic frequencies of the problem in which the field satisfies dissipative boundary conditions, ingoing at the event horizon and outgoing at infinity, i.e.,
        \begin{equation}
            \psi(x)\sim\begin{cases}
			e^{-i\,\omega\,x}, \text{ for } x\rightarrow -\infty, \\
			e^{i \,\omega x}, \text{ for } x\rightarrow \infty.
		\end{cases}
        \end{equation}
        
        The quasinormal frequencies were obtained using the hyperboloidal framework and a modified Leaver method as presented in the Appendix \ref{AP:MethodsToCalculateFS}.
        We follow the definitions present in \cite{Cheung:2021bol}, where we identify $\varpi$ as the fundamental mode frequency of the perturbed potential $V_{\Gamma}$ and $\omega^{(\Gamma)}_{n}$ are the quasinormal frequencies that deform continuously from the original unperturbed modes $\omega^{(0)}_{n}$, with $\omega^{(\Gamma)}\equiv \omega^{(\Gamma)}_{0}$ being the continuous deformation from the fundamental mode, $\omega^{0}\equiv\omega^{0}_{0}$. We measure the variation $\Delta\omega^{(\Gamma)}_{n}=\omega^{(\Gamma)}_{n}-\omega^{(0)}_{n}$.

        The quasinormal frequencies react to the perturbation at different stages. In the case of a general small perturbation on the potential, the position of perturbation and the imaginary part of the modes from the unperturbed case dictate when the spectrum is modified, for a small disturbance on the potential the modes vary as \cite{Leung:1999iq, Cardoso:2024mrw}
        \begin{equation}\label{EQ:variationOmega}
            \delta\omega\sim \epsilon e^{-2\,r_{0} \omega_{I}},
        \end{equation}
        where $r_{0}$ locate the disturbance, $\omega_{I}$ is the imaginary part of the quasinormal frequency and $\epsilon$ is the magnitude of the perturbation. This disturbance will be in a perturbation regime as long we have $\epsilon\ll e^{2\,r_{0}\omega_{I}}$, which means the modes will suffer small variations. In our case, it is the vortex properties and the circulation at infinity that dictate the range of this regime, since we have
        \begin{equation}
         \epsilon = \left(\frac{\Gamma}{2\pi\sigma_{v}^{2}}\right)^{2}.
        \end{equation}

\section{Results}\label{sec:results}

We consider as the main setup to start the analysis the following values of the flow properties: $\mu_{\rm k} = 8.917~\times~10^{-7}\,\, {\rm m}^2/{\rm s}$, $\alpha = 1.7834~\times~10^{-6} \,\, {\rm s}^{-1}$, $\sigma_{v}=\sqrt{\mu_{k}/\alpha}=2^{-1/2}$,  $\Gamma = 0.1\,\, {\rm m}^2/{\rm s}$, and $c=1\,{\rm m}/{\rm s}^{2}$. 

    \begin{figure*}[htpb!]
    \centering
     \includegraphics[width=.45\textwidth]{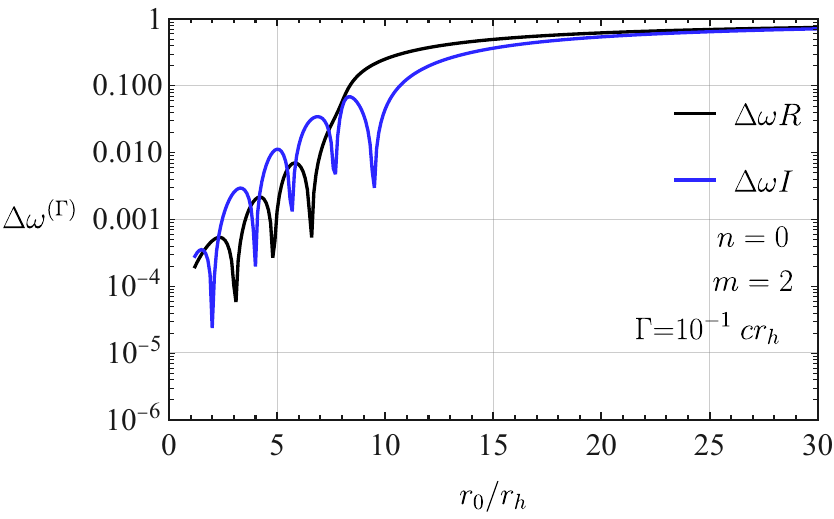}
     \includegraphics[width=.45\textwidth]{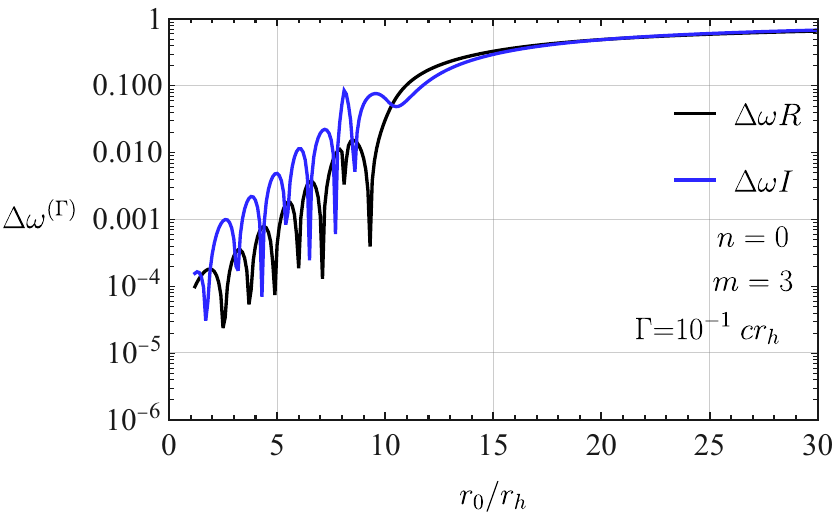}
     
      \caption{The migration distance in the real and imaginary part of the frequency of the fundamental mode for $m=2$ (left) and $m=3$ (right), with $\Gamma=10^{-1}m^{2}/s$.}
     \label{FIG:VarFundModeAndOTforGamma01}
     \end{figure*}
     \begin{figure*}
          \includegraphics[width=.45\textwidth]{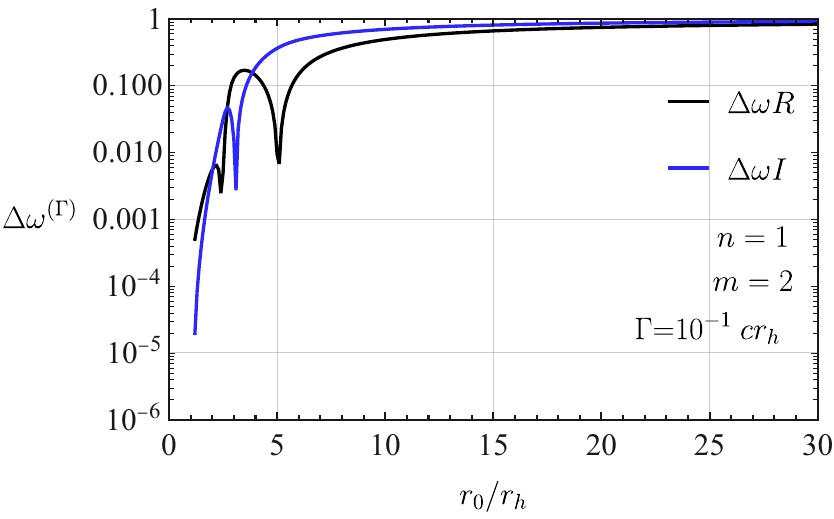}
     \includegraphics[width=.45\textwidth]{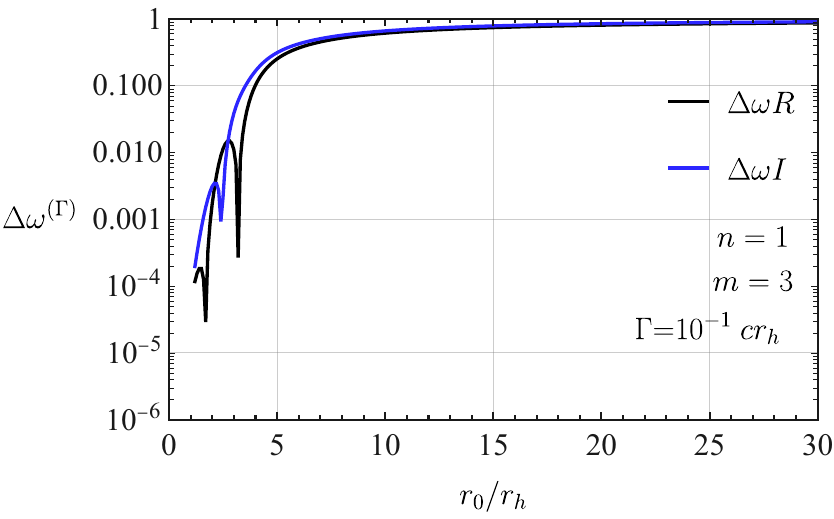}
     \caption{The migration distance in the real and imaginary part of the frequency of the first overtone, the for $m=2$ (left) and $m=3$ (right), with $\Gamma=10^{-1}m^{2}/s$.}
     \label{FIG:VarOvertonAndOTforGamma01}
     \end{figure*}
    
        The overtones having a greater imaginary part are more sensible to the portential perturbation. In Fig. \ref{FIG:VarFundModeAndOTforGamma01} and \ref{FIG:VarOvertonAndOTforGamma01}, we see the change in the fundamental mode and the first overtone as we track it from the unperturbed case, considering both $m=2,\,3$, and as the position of the vortex $r_0$ increases. The first overtones present a faster and longer migration compared to the fundamental case.

\begin{figure*}[htpb!]
\centering
\includegraphics[width=.45\textwidth]{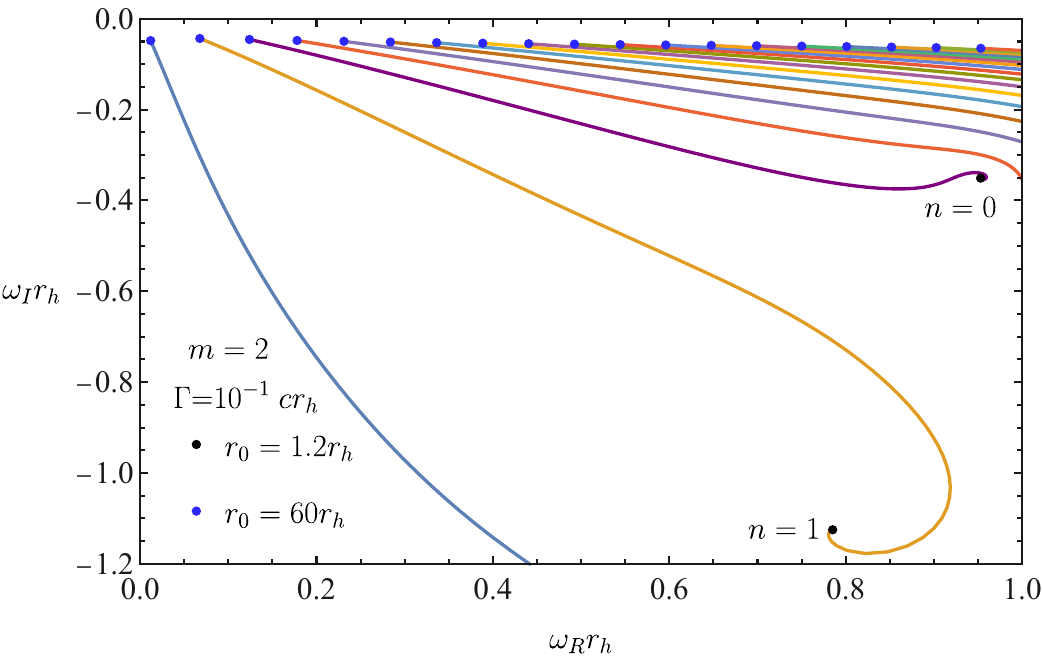}
\includegraphics[width=.45\textwidth]{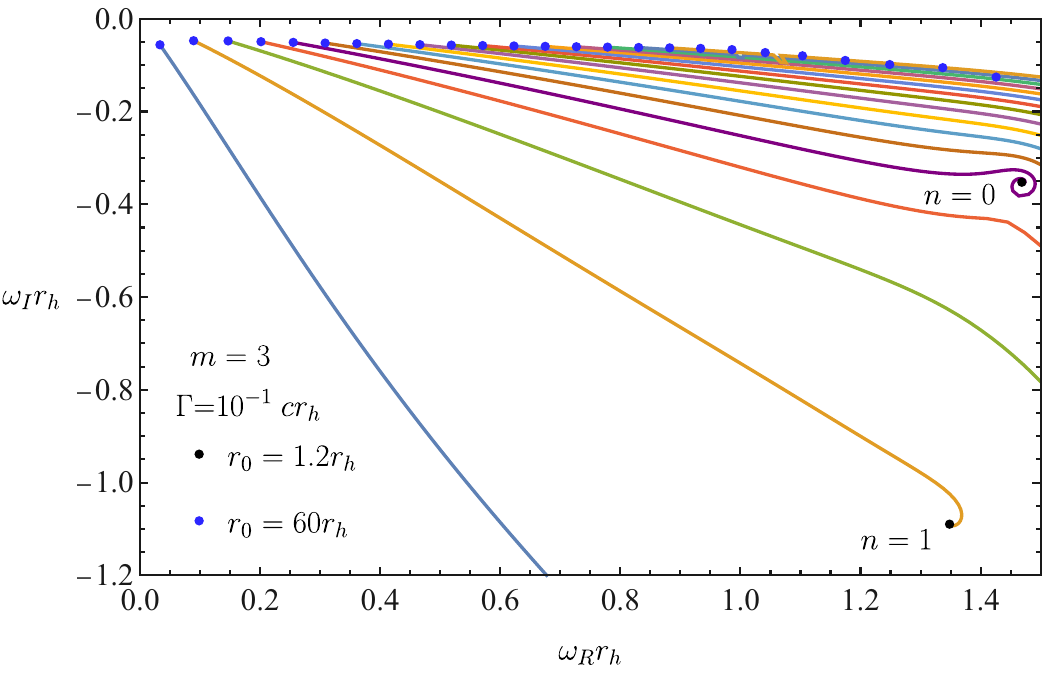}

\caption{The migration of the fundamental, the first overtone, and of the new modes that appear due to the perturbation, as we move the position of the vortex.}
\label{FIG:TrackOfModesGamma01}
\end{figure*}

        Including the vortex changes the original quasinormal frequencies and it may even add new modes to the spectrum. The existence of new modes can be predicted as we are adding another potential barrier where additional quasinormal modes can be trapped. In Fig.~\ref{FIG:TrackOfModesGamma01}, we see the migration of the unperturbed modes and the additional modes. The spectrum starts to exhibit modes with similar imaginary and equally spaced real parts, which indicates that we might have ``echoes'' in the ringdown due to trapped modes between the barriers~\cite{Cheung:2021bol,Jaramillo:2021tmt}.
        
\begin{figure}[htpb!]
\centering
\includegraphics[width=.48\textwidth]{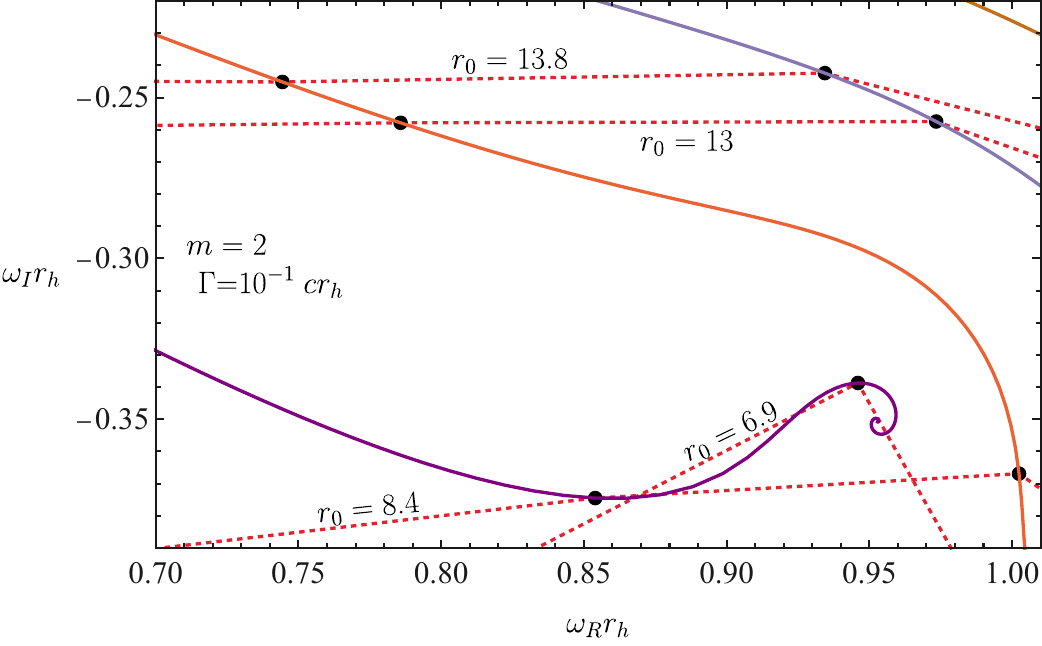}

\caption{The change in the fundamental quasinormal mode (overtaking) as we vary the position of the vortex. The first change happens at $r_{0}=8.4 r_{h}$ when the fundamental mode ceases to be the one inherited from the unperturbed case. At $r_{0}=13.8 r_{h}$ the fundamental mode jumps to values where $\varpi_{R}<0.3$.}
\label{FIG:Overtaking2}
\end{figure}

        The overtones migrate faster than the fundamental mode, they can even take the position of fundamental modes\footnote{Here we recall that the fundamental mode is characterized by being the one that lives longer, having the smallest $-\omega_i$. Therefore, the hyerarchy of the modes is obtained by analyzing their imaginary parts.}, presenting a lower imaginary part than the original fundamental mode, hence \textit{overtaking} its position. In Fig.~\ref{FIG:Overtaking2}, we see the position of the vortex in which discontinuous jumps occur in the fundamental mode.  Until $r_{0}=8.4r_{h}$ the changes in the fundamental mode are in the perturbative regime, Eq.~(\ref{EQ:variationOmega}), hence it is the same as the unperturbed case, $\omega^{\Gamma}_{0}=\varpi$ after this $\varpi$ jumps to the next higher real part mode. There is a further jump at $r_{0}=13.8 r_{h}$ where it jumps for smaller values for the real part, and in this configuration, it coincides with the migrated unperturbed fundamental mode.

\begin{figure}[th]
        \includegraphics[width=.48\textwidth]{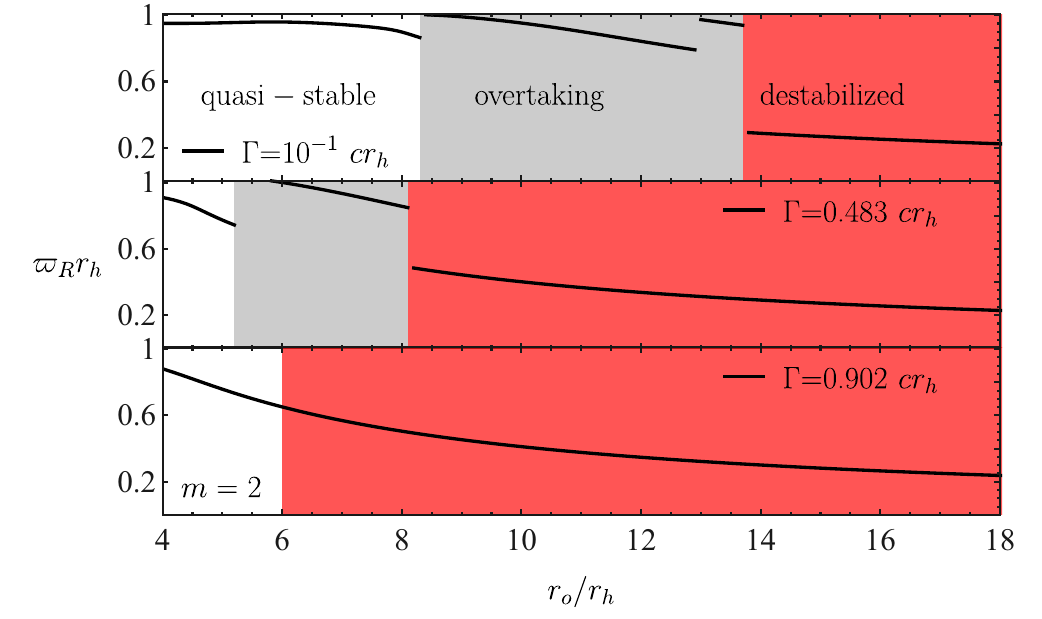}\qquad
        \caption{The real part of the fundamental mode, $\varpi_{R}$, for three values of $\Gamma$, and varying the position of the vortex. The overtaking of the fundamental mode in gray, and its absence for higher values of $\Gamma$.}
        \label{FIG:VarpiDiffGammas}
    \end{figure}

        The migration of the unperturbed fundamental mode is slower compared to all the other modes, and it starts to present significant deviations from the unperturbed values when the vortex surpasses the condition (\ref{EQ:variationOmega}). In Fig.~\ref{FIG:VarpiDiffGammas}, we see the variation in the real part of the fundamental mode as we vary the position of the vortex, for three different values of $\Gamma$. As we increase the influence of the vortex, we see that the destabilization by overtaking vanishes, and the fundamental mode for a given vortex is the one migrating from the unperturbed case leading to a destabilization only by migration. In order words, depending on the amplitude of the perturbation, here the vortex, we might have a frequency spectrum in which the fundamental mode migrates but does not jump between different modes, a consequence from trespassing the condition (\ref{EQ:variationOmega}) even for small values of $r_0$. This feature is not reported in the parameter range analyzed in Ref.~\cite{Cheung:2021bol}. However,  Fig. $2$ in Ref.~\cite{Cheung:2021bol} seems to indicate that the overtaking region is decreasing as for higher $\epsilon$, so that one may find a similar regime if the parameter range is enlarged.

\begin{figure}
    \centering
    \includegraphics[width=.48\textwidth]{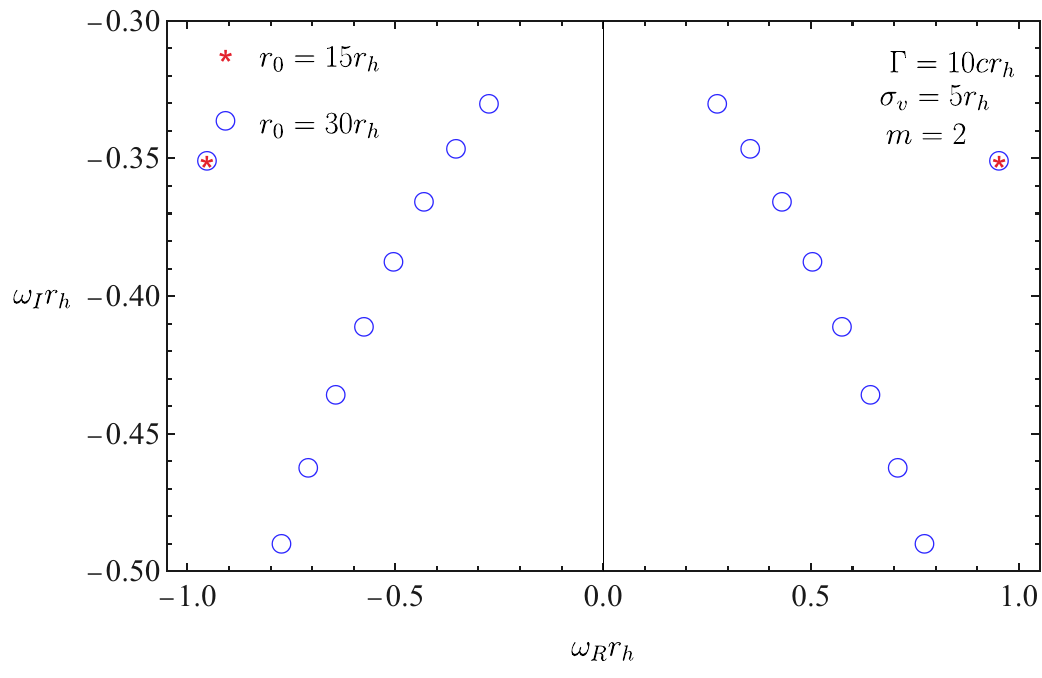}
    \caption{Quasinormal frequencies in the case of a vortex with $\sigma_{v}=5\,r_{h}$, $\Gamma=10 c r_{h}$ and for two location of the vortex $r_{0}=15,\,30 \,r_{h}$. New modes for this vortex's sparse configuration resemble massive scalar field spectrum in de Sitter spacetime.}
    \label{FIG:QNM_SigmaV5Varr0}
\end{figure}

         We can also vary the vortex width, $\sigma_{v}$, which spreads the influence of the vortex. In the frequency domain, we see a similar feature for the migration of the modes, but with more modes being allocated between the fundamental mode and the vertical axis. In Fig.~\ref{FIG:QNM_SigmaV5Varr0}, we plot the quasinormal frequencies found for the configuration with $\sigma_{v}=5 r_{h}$ and $\Gamma=10 c r_{h}$, this type of spectrum resembles the one for a massive scalar field in a Schwarschild-de Sitter spacetime, where the de Sitter modes acquire a real part \cite{Correa:2024xki}. For the vortex at $r_{0}=30\, r_{h}$ the fundamental mode is not yet in a long migration path, but the additional modes have already destabilized the spectrum.
         
        The frequency and time domain present different responses to the presence of the vortex.  The frequency spectrum is very sensible to small perturbations, for small values of $r_{0}$, the modes already present some migration displacements, but the fundamental mode is almost the same. As the location of the vortex is increased, the whole spectrum is modified presenting a similar behavior to that describing ``echoes'' and trapped modes due to the two potential barriers. 
        
\begin{figure*}[htpb!]
\centering
\includegraphics[width=0.49\textwidth]{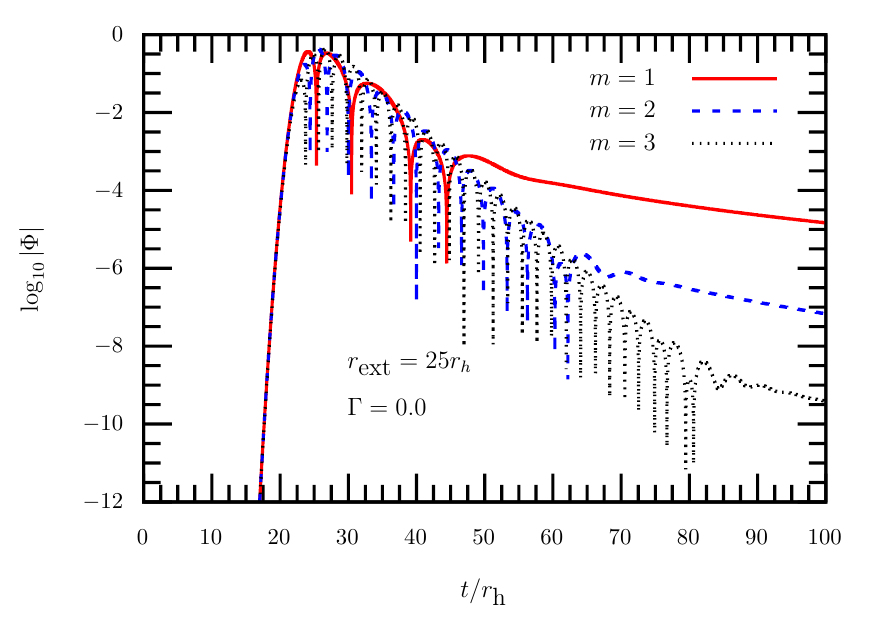}\includegraphics[width=0.49\textwidth]{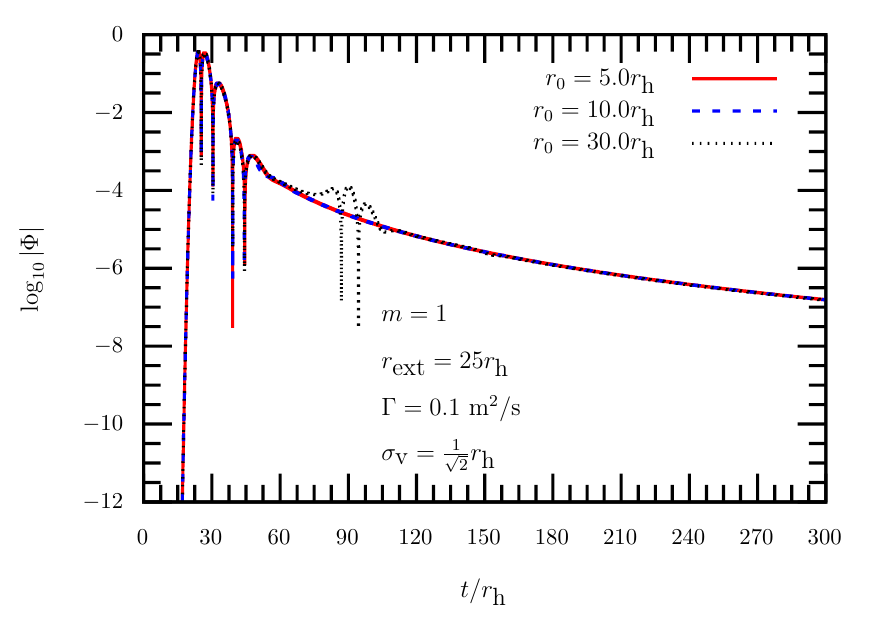}
\includegraphics[width=0.49\textwidth]{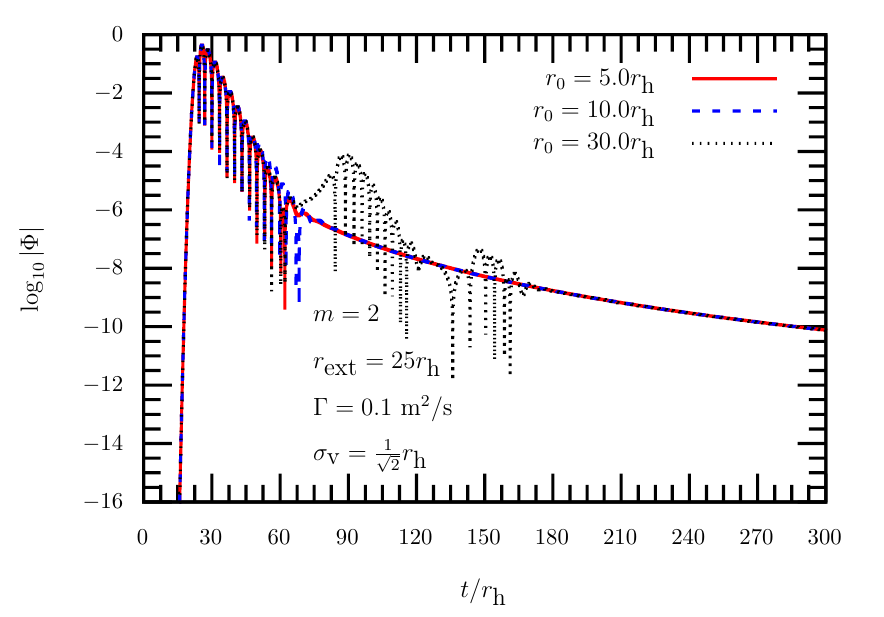}\includegraphics[width=0.49\textwidth]{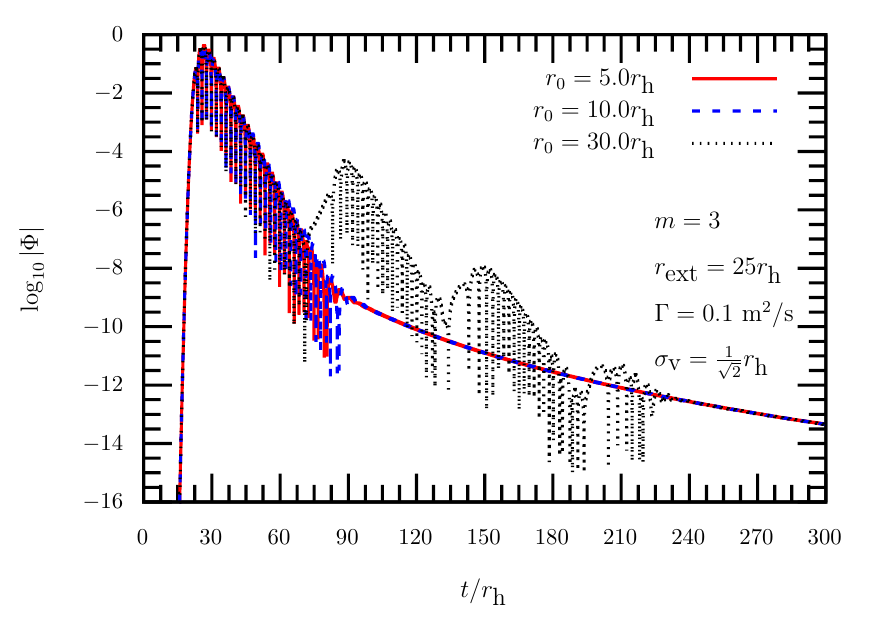}

\caption{Time-domain profiles for azimuthal number $m = 1,\,2,\, 3$. The wave is extracted at $r_{\rm ext} = 25\,r_{\rm h}$. The ringdown is almost the same, the vortex influence is stronger for higher values of $r_{0}$ and includes echoes at intermediate times.}
\label{Ring}
\end{figure*}

        In the time domain, the ringdown is directly related to the fundamental mode and potential height while the time tails are associated with the backscattering with the curvature at long distances. The addition of the vortex at small values of $r_{0}$ presents additional oscillations and echoes due to the reflections and interactions between the barriers of the potential, changing the end of the ringdown phase and the start of late-time tails, but barely modifying the fundamental mode as we can see on Fig.~\ref{Ring}. A discussion relating the causal structure of the Green function for this type of system and the signals in the tail is presented in Ref.~\cite{PhysRevD.110.084018}. At higher values of $r_{0}$ the response in the time domain is concentrated in intermediate times presenting signals similar to echoes, trapped modes, and long-lived ringing depending on the width of the vortex, Fig.~\ref{Ring}. The ringdown phase is barely modified, hence the fundamental mode through the time domain does not present a significant destabilization as seen in the frequency domain, which we can see comparing the $\omega^{0}$ for $r_{0}=15 r_{h}$ as obtained through the time and frequency domain in Table~\ref{table2}. 
        
\begin{figure*}
    \centering
    \includegraphics[width=0.49\textwidth]{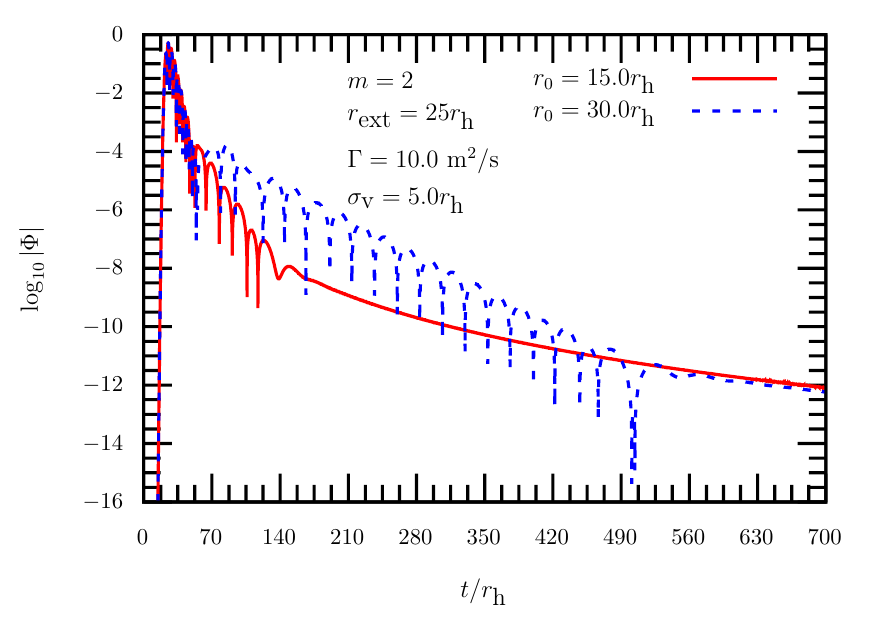}\includegraphics[width=0.49\textwidth]{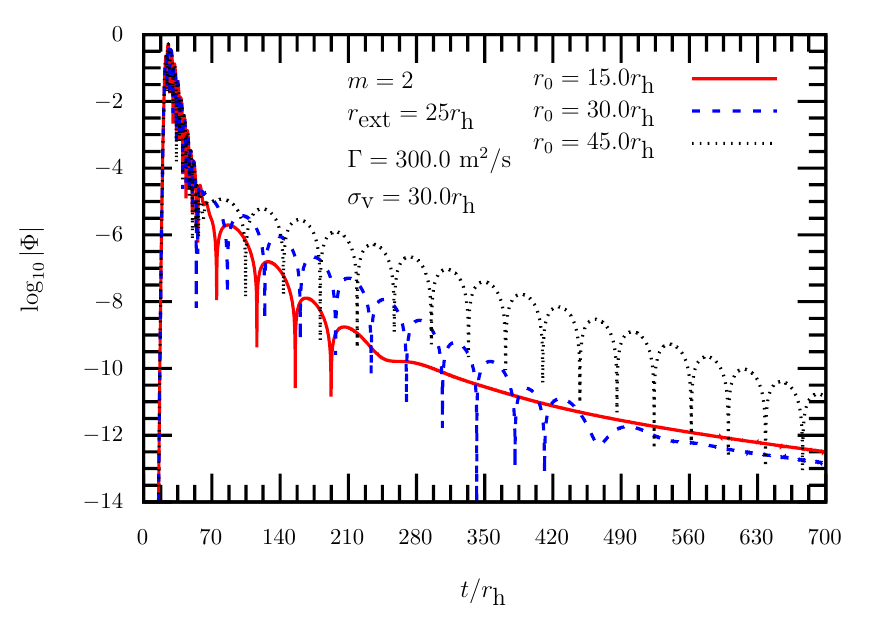}
    \caption{Time-domain profile for considering potentials with high values of $\sigma$. We see that a higher width $\sigma_v$ generates lower frequency additional modes, living longer than the original ones.}
    \label{fig:Ring2}
\end{figure*}

        {In the time domain, Fig.~\ref{Ring}, we see in the top left panel the decay of a perturbation in the system without vortex and for different values of $m$. In the other panels, we fix the value of $m$ in each panel and vary the position of the vortex, and for these configurations, the imprint of the vortex in the time evolution introduces some trapped modes or echoes due to sequential scattering between the potential barrier and the vortex. In Fig.~\ref{fig:Ring2}, we see the imprinting of a more sparse vortex, which introduces oscillations partially changing the ringdown and some intermediate times, until it gets dominated by the tail. The additional ringing in this later case resembles the decay of massive fields in black hole spacetimes, which are related to the quasibound states in such systems \cite{Dolan:2007mj} or even the de Sitter modes in Schwarzschild-de Sitter spacetime~\cite{Correa:2024xki}. These curious behavior opens the possibility of mimicking additional features of fields around black holes in analogue systems.}

\begin{table*}
\centering \caption{Quasinormal frequencies of the fundamental mode for different azimuthal numbers $m$ and for $\Gamma = 0.1 c r_{h}$ and without the vortex, obtained through the frequency domain approach and the time-domain profiles extracted from Eqs.~\eqref{ode1} and~\eqref{ode2}.}

\begin{tabular}{@{}cccc@{}}
\hline \hline

\multicolumn{1}{c}{} & \multicolumn{3}{c}{$\omega$}\\
\hline 

\multicolumn{4}{c}{Time domain} \\
\hline 
\multicolumn{1}{c}{$m$} & \multicolumn{1}{c}{\textit{No vortex}} & \multicolumn{1}{c}{$r_0 = 10.0\,r_{\rm h}$} &\multicolumn{1}{c}{$r_0 = 15.0\,r_{\rm h}$}\\ \hline

2   & \,\,  $  0.952658 - 0.357277 \,i $   & \,\, $0.963361 - 0.338284  \,i $     & \,\, $ 0.956752 - 0.355861 \,i $           \\

3   & \,\, $ 1.46727 - 0.352396\,i $  & \,\,  $ 1.45968 - 0.345949 \,i $     & \,\,  $ 1.46647 - 0.353332 \,i $           \\
\hline 
\multicolumn{4}{c}{Frequency domain} \\
\hline
2 & $ 0.95273-0.35074 i$ & $0.95277-0.29940  i$& $0.87625 - 0.22557 i$\\
3 & $1.46854-0.35243 i$ & $1.42351-0.32989 i$& $0.98989-0.24918 i$\\
\hline \hline
\end{tabular}
\label{table2}
\end{table*}

	\section{Conclusion}\label{sec:conclusion}
We have studied how small disturbances in the wave equation modelling an analogue black hole spacetime modify the QNM spectrum and the time-domain evolution of this configuration. Similar to what occurs in black hole spacetimes, we consider QNM instability resulting from an ``elephant and the flea” setup, where a small bump is introduced in the wave equation’s effective potential. Most importantly, in the draining bathtub model, such modifications arise naturally from the Navier-Stokes equation, with the small bump directly interpreted as a vortex in the fluid flow. This top-down derivation provides a more physically grounded framework to analyze how the properties of this disturbance affect the underlying simulated geometry.

The impact of the perturbation depends on the domain of analysis. In the frequency domain, the system is highly sensitive to deviations from the unperturbed case. Overtones are more prone to migration, while the fundamental modes exhibit greater stability under such disturbances. As the vortex moves farther from the center, the fundamental mode becomes increasingly destabilized, migrating significantly from its original value. However, in certain parameter ranges, the fundamental mode remains the same despite migrating—suggesting that it is perturbed but not necessarily overtaken. This feature was not report in Ref.~\cite{Cheung:2021bol}, as the range of $\epsilon$ analyzed there was insufficient to reveal the absence of an overtaking region. While the difference in behavior may stem from fundamental differences between acoustic black holes and Schwarzschild black holes, the results from Ref.~\cite{Cheung:2021bol} show a tendency of a decreasing overtaking region as the bump's height increases.

The time-domain response, on the other hand, exhibits a distinct behavior. A modification in the potential near its peak leaves visible imprints on the ringdown phase, whereas a perturbation farther away primarily affects the late-time tail of the signal. Due to the small size of the vortex, the ringdown is only marginally affected, meaning that the fundamental mode extracted from this phase remains largely unchanged compared to the significant shifts observed in the frequency domain. However, the tail of the signal exhibits clear signatures of the perturbation, including echoes, trapped modes, and long-lived oscillations, depending on the characteristics of the vortex. A more localized vortex leads to stronger trapped modes and potential echoes, while a more diffuse vortex produces oscillations reminiscent of massive fields in black hole spacetimes.

These results extend the study of fundamental mode destabilization from black hole spacetimes to analogue systems, with the \textit{bump} interpreted as a Burgers-Rott vortex~\cite{BURGERS1948171,Rott1958ZaMP}. The experimental realization of these effective spacetimes has already enabled the detection of quasinormal modes, their enhancement in confined systems, and superradiance in scattering waves \cite{Torres:2016iee,Torres:2020tzs,Smaniotto:2025hqm}. It is reasonable to expect that real experimental setups will not be perfectly ``clean,’’ making it crucial to understand how impurities and background perturbations influence the spectra. Therefore, our results—along with recent pseudospectral analyses of rotating black hole analogues~\cite{Tobias2025}—serve as an important step toward extending the study of quasinormal mode spectral instability from astrophysical black holes to laboratory analogues.

	\begin{acknowledgments}
    MMC would like to thank the Strong Group at the Niels Bohr Institute (NBI) for their kind hospitality during the final stages of this work. The Tycho supercomputer hosted at the SCIENCE HPC center at the University of Copenhagen was used for supporting this work.
    CFBM, MMC, and LAO acknowledge Fundação
    Amazônia de Amparo a Estudos e Pesquisas (FAPESPA),
    Conselho Nacional de Desenvolvimento Científico e Tecnológico (CNPq) and Coordenação de Aperfeiçoamento de 
    Pessoal de Nível Superior (CAPES) – Finance Code 001,
    from Brazil, for partial financial support.
RPM acknowledges support from the Villum Investigator program supported by the VILLUM Foundation (grant no. VIL37766) and the DNRF Chair program (grant no. DNRF162) by the Danish National Research Foundation. 
The Center of Gravity is a Center of Excellence funded by the Danish National Research Foundation under grant No. 184.
	\end{acknowledgments}
	
	\appendix
	
	\section{Methods to compute quasinormal modes}\label{AP:MethodsToCalculateFS}

 \subsection{Hyperboloidal method}
     To obtain the frequency spectrum we use the hyperboloidal framework together via the scri-fixing and a spatial compactification \cite{Zenginoglu:2007jw,Zenginoglu:2011jz,PanossoMacedo:2023qzp, PanossoMacedo:2024nkw} where we make a coordinate transformation from ($t$, $r$) to ($\tau$, $\sigma$), given by:
        \begin{equation}
        t=\tau-H(\sigma), \hspace{0.5 cm} r=\frac{r_{h}}{\sigma},
        \end{equation}
        with $\sigma \in [0,1]$, with $1$ the event horizon and $0$ the infinity, and the ``height'' function $H(\sigma)$ is constructed from the singular and regular parts of the tortoise coordinate, $x(\sigma)\equiv x(r(\sigma))$, cf. Eq.~(\ref{EQ:tortoise}), such that we work in the minimal gauge \cite{Ansorg:2016ztf,PanossoMacedo:2018hab,PanossoMacedo:2019npm,PanossoMacedo:2023qzp}. Assuming a Fourier decomposition $\Phi=e^{-i \omega t}\psi(r)$ in Eq.~(\ref{waveeq3}), and within the hyperboloidal framework we reach the differential equation
        \begin{equation}\label{EQ:waveEqSigma}
            \begin{aligned}
            \frac{d}{d\sigma}\left(p(\sigma)\frac{d\psi(\sigma)}{d\sigma}\right)+&\left(\omega^{2}w(\sigma)-\frac{V(\sigma)}{p(\sigma)}\right)\psi(\sigma)+ \\
            -&i\omega \left(2\gamma(\sigma)\frac{d\psi(\sigma)}{d\sigma}+\gamma'(\sigma)\psi(\sigma)\right)=0,
            \end{aligned}
        \end{equation}  
        with
        \begin{equation}
            \begin{aligned}
            p(\sigma)=-\frac{1}{x'(\sigma)},\\
            \hspace{0.2 cm} \gamma(\sigma)=\frac{dH}{d\sigma}p(\sigma), \\
            w(\sigma)=\frac{1-\gamma(\sigma)^2}{p(\sigma)}.
            \end{aligned}
        \end{equation}
       The boundary conditions, only waves leaving the system, are naturally satisfied when working in the hyperboloidal framework. We can perform a first-order reduction in time $\bar{\phi}=d\psi/d\tau=-i \omega\,\psi$, leading us to the system of equations
       \begin{equation}	
		\mathbf{L}\vec{U} =-i\omega \vec{U},
        \end{equation}
        with
        \begin{equation}
            \mathbf{L}=\left(\begin{matrix}
			0 & 1 \\
			w(\sigma)^{-1}L_{1} &	\, w(\sigma)^{-1}L_{2}\\
		\end{matrix}\right),\hspace{0.5 cm}
          \vec{U}= \left(\begin{matrix}
			\psi  \\
			\bar{\phi}\\
		\end{matrix}\right),
        \end{equation}
        and the components of the $\mathbf{L}$ operator given by
        \begin{equation}
        \begin{aligned}
            L_{1}=\frac{d}{d\sigma}\left(p(\sigma)\frac{d}{d\sigma}\right)-\frac{V(\sigma)}{p(\sigma)},\\
            L_{2}=2 \gamma(\sigma)\frac{d}{d\sigma}+\gamma'(\sigma).
            \end{aligned}
        \end{equation}        
        To find the eigenvalues of the operator $\mathbf{L}$, we use the spectral methods where we expand our functions in a set of basis functions as
        \begin{equation}
            \psi(\sigma)=\sum_{i=0}^{N}c_{i}T_{i}(\sigma).
        \end{equation}
        where we use the Chebyshev polynomials, given by $T_{j}(\xi)=\cos(j\,\arccos \,\xi)$. We use the Chebyshev-Lobatto grid for the collocation points to reach a system of $N+1$ equations, from which we find the quasinormal frequencies as the system's eigenvalues. 
 \subsection{Modified Leaver's method}

 The Leaver method consists in performing a Frobenius expansion of the solution in a way that it satisfies both boundary conditions. The coefficients in this expansion can be obtained through a recurrence relation, leading to a continued fraction. In our case, due to the form of $\Omega$, it is not trivial to get a series solution covering the whole space. However, by noticing that the problematic part of the modes comes from the divergence of the radial solutions at numerical infinity, we can pose a series solution in a region where $\Omega\approx 0$ in the large $r$ region, similarly to what is done for relativistic stars~\cite{Benhar:1998au,Pani:2013pma,Macedo:2016wgh}. Since the bump decays much faster than the background potential, the approximation is justifiable. 

In a region where $\Omega\approx 0$, in the right-hand side of the potential bump, we can make the following expansion
\begin{equation}
    \Phi(r)=\left(\frac{r-r_h}{r+r_h}\right)^{i\omega r_h/2}e^{i\omega r}\psi(v(r)),
\end{equation}
with $v(r)=1-a/r$, where $a$ is a radial distance outside the vortex region. By using the above expression, we obtain the following equation
\begin{align}
 &(v-1)^2 (a+r_h (v-1)) (a-r_h v+r_h)\frac{d^2\psi}{dv^2}+\nonumber\\
&2 i a^3 \omega +2 a^2 (v-1)-4 r_h^2 (v-1)^3\frac{d\psi}{dv}+\nonumber\\
&\frac{1}{4} \left(a^2 \left(1-4 m^2\right)-5 r_h^2 (v-1)^2\right)\psi=0.\label{eq:veq},
\end{align}
We now search for a solution of the form
\begin{equation}
    \psi=\sum_{n=0}^{\infty}b_n v^n,
\end{equation}
and using Eq.~\eqref{eq:veq} we find the following five-term recurrence relation
\begin{align}
\alpha_n b_{n+1}+\beta_n b_{n}+\gamma_n b_{n-1}+\delta_n b_{n-2}+\nonumber\\
\sigma_n b_{n-3}=0,~n\geq 3
\end{align}	
with
\begin{align}
&\alpha_n=n (n+1) (a-r_h) (a+r_h),\\
&\beta_n=2 n \left(i a^3 \omega -a^2 n+2 n r_h^2\right),\\
&\gamma_n=\frac{1}{4} \left[a^2 \left((1-2 n)^2-4 m^2\right)\right.\nonumber\\
&\left.+(-24 (n-1) n-5) r_h^2\right],\\
&\delta_n=\frac{1}{2} [8 (n-2) n+5] r_h^2,\\
&\sigma_n=-\frac{1}{4} [4 (n-3) n+5] r_h^2.
\end{align}
The above recurrence relation can be solved to find the function $\psi(v)$ [and therefore $\Phi(r)$] outside the vortex region. Notice that this solution is determined up to a constant: all coefficients $a_n$ are written in terms of $b_0$, which we can set to unity since the equation is linear. We also guarantee that the solution is convergent by selecting the minimal solution~\cite{Benhar:1998au}. Let us call the solution constructed from the recurrence relation $\Phi^+(r)$. This solution satisfies the required boundary condition for quasinormal modes at infinity.

To ensure that we have the required conditions on the inner part of the vortex, we need to construct another solution, say $\Phi^-$, that represents ingoing waves at the horizon. We start with the boundary condition at the horizon
\begin{equation}
    \Phi(r\approx r_h)\approx e^{-i\omega x}\sum_{j=0}^{N}c_j(r-r_h)^j,
\end{equation}
where $N$ depends on the numerical accuracy required for the boundary condition, and the coefficients $c_j$ are found by recursively solving the differential equation expanded at infinity. We use the above boundary condition and integrate outwards to a point $a$, over the region where $\Omega$ has its influence. The point $r_m$ is located at the rightmost part of the radial distance, where $\Omega\approx0$. In this way, we construct a solution $\Phi^-$ at $r=a$.

Finally, to ensure that both conditions at infinity are satisfied, we impose that the Wronskian of the two vanishes at $r=a$, i.e.,
\begin{equation}
    \Phi^+(a)\frac{d\Phi^-(a)}{dr}-\Phi^-(a)\frac{d\Phi^+(a)}{dr}=0.
    \label{eq:wronskian}
\end{equation}
Eq.~\eqref{eq:wronskian} is only satisfied for a given set of frequencies $\omega$ which are the quasinormal frequencies of the system.

To validate our results, in Table \ref{tab:mode_comp} we compare the quasinormal frequencies computed through the hyperboloidal and Leaver approaches, in the Burgers-Rott vortex case.
\begin{table*}
    \centering
    \caption{Quasinormal frequencies computed through the Hyperboloidal and Leaver procedures. Here we consider the Burgers-Rott vortex, with $m=2$ and $\Gamma=10^{-1}cr_{h}$ as in the left panel of Fig. \ref{FIG:TrackOfModesGamma01}.}
    \label{tab:mode_comp}
    \begin{tabular}{l |l |l}
    \hline
    $r_0$         & Hyperboloidal & Leaver \\\hline
    $7 r_h$       & $0.9435478775-0.3389347899 i$& $0.9435478776 - 0.3389347900 i$\\
                  & $0.5725367526-0.4977815967 i $& $0.5725367526 - 0.4977815966  i$\\
                  & $0.1316048468 -0.5447875576  i $& $0.1315999072 - 0.5447838413 i$\\
                  & $1.0886810442-0.5639314843  i $& $1.0886810453 - 0.5639314820 i$\\
      \hline
    $10 r_h$       & $0.9527663590-0.2993980903  i$&$0.9527663590 - 0.2993980903 i$\\
                   & $0.7152705271 - 0.3351820365 i $ &$0.7152705271 - 0.3351820365 i$\\
                   & $0.4012226878 - 0.3439356626 i $ &$0.4012226878 - 0.3439356626 i$\\
                   &$0.0836834097-0.3695150455  i$ &$0.0836837479 - 0.3695143347 i$\\
      \hline
    $15 r_h$     & $0.8762555041 - 0.2255769994 i$ &$0.8762555041 - 0.2255769991 i$\\
                 &$0.2687668292 - 0.2206412271 i$ &$0.2687668292 - 0.2206412271 i$\\
                 & $0.4850567508 - 0.2235448975 i$&$0.4850567508 - 0.2235448975 i$\\
                 & $0.6891553432 - 0.2268040844 i$&$0.6891553432 - 0.2268040845 i$\\
      \hline
    \end{tabular}
\end{table*}

	\bibliography{SpectralnstabilityABH}
\end{document}